\documentclass[12pt,british,epsfig,a4paper,here,amssymb]{article}
\usepackage[T1]{fontenc}
\usepackage[latin1]{inputenc}
\usepackage{float}
\usepackage{amstext}
\usepackage{amssymb}
\usepackage{mcite}
\usepackage{euscript}
\usepackage{calc}
\usepackage{babel}

\makeatletter

\newif\ifpdf 
\ifx\pdfoutput\undefined 
\pdffalse 
\else 
\pdfoutput=1 
\pdftrue 
\fi

\newcommand{\PRnum}     {CERN-PH-EP/2004-041}
\newcommand{\Date}      {12 August, 2004}
\newcommand{\Author}    {The OPAL Collaboration}

\ifpdf
\usepackage{thumbpdf}
\usepackage[pdftex,colorlinks=true,
                   pdfstartview=FitV,
                   linkcolor=blue,
                   citecolor=blue,
                   urlcolor=blue,
                   ]{hyperref}
\usepackage{pslatex}
\else
\newcommand{\href}[2]{#2}
\fi

\usepackage{graphicx}

\newcommand{\RSMhLimitMinObs}{58\GeV}
\newcommand{\RSMhLimitMinExp}{54\GeV}
\newcommand{\RSMrSensitivLambdaMax}{0.8\TeV}
\newcommand{\RSMScanStepLow}{3\GeV}
\newcommand{\RSMScanStepHigh}{30\GeV}

\newcommand{\sqrts}{\sqrt{\mathrm{s}}}
\newcommand{\deq}{=}

\newcommand{\hzero}{\mathrm{h}}
\newcommand{\Hsm}{\mathrm{H}_{\mathrm{SM}}}
\newcommand{\mHsm}{m_{\Hsm}}

\newcommand{\mW}{m_{\mathrm{W}}}
                  
\newcommand{\tautau}{\tau^{+}\tau^{-}}
\newcommand{\ee}{\mathrm{e}^{+}\mathrm{e}^{-}}
\newcommand{\qq}{\mathrm{q}\overline{\mathrm{q}}}
\newcommand{\bb}{\mathrm{b}\overline{\mathrm{b}}}
\newcommand{\glgl}{gg}
\newcommand{\ra}{\rightarrow}
      
\newcommand{\Zzero}{\mathrm{Z}}

\newcommand{\invpb}{\, \mathrm{pb}^{-1}}
\newcommand{\mum}{\,\mu\mathrm{m}}

\newcommand{\gev}{\,\mathrm{Ge}\kern-0.08em  \mathrm{V}}
\newcommand{\GeV}{\, \mathrm{Ge}\kern-0.08em  \mathrm{V}}
\newcommand{\TeV}{\, \mathrm{Te}\kern-0.08em  \mathrm{V}}
\newcommand{\MeV}{\,\mathrm{Me}\kern-0.08em  \mathrm{V}}

\newcommand{\MrhD}{\left(m_{+}^{2}-m_{-}^{2}\right)^{2}-\frac{144\xi^{2}\gamma^{2}}{Z^{2}}m_{+}^{2}m_{-}^{2}}
\newcommand{\MrhQ}{(m_{+}^{2}+m_{-}^{2})}
\newcommand{\RSmr}{m_{\mathrm{r}}}
\newcommand{\RSmh}{m_{\mathrm{h}}}
\newcommand{\RSFundmr}{\widetilde{m}_{\mathrm{r}}}
\newcommand{\RSFundmh}{\widetilde{m}_{\mathrm{h}}}
\newcommand{\RSEigenr}{\hat{\mathrm{r}}}
\newcommand{\RSEigenh}{\hat{\mathrm{h}}}
\newcommand{\RSFundr}{\tilde{r}}
\newcommand{\RSFundh}{\tilde{h}}
\newcommand{\RSRadion}{\mathrm{r}}
\newcommand{\RSHiggs}{\mathrm{h}}
\newcommand{\MPlanck}{M_{\mathrm{Pl}}}
\newcommand{\RSScale}{\Lambda_{\mathrm{W}}}
\newcommand{\VEV}{v}

\newcommand{\PlanckScale}{10^{15}\TeV}

\makeatother

\begin{document}
\begin{titlepage}

\centerline{{\large EUROPEAN ORGANIZATION FOR NUCLEAR RESEARCH}} \bigskip

\begin{flushright}\PRnum\\
 \Date\end{flushright}

\bigskip\bigskip\bigskip

\begin{center}\textbf{\Large Search for Radions at LEP2}\end{center}{\Large \par}

\bigskip

\begin{center}{\Large\Author{}}\end{center}\par

\bigskip

\begin{center}{\large Abstract}\end{center}{\large \par}

A new scalar resonance, called the radion, with couplings to fermions
and bosons similar to those of the Higgs boson, is predicted in the
framework of Randall-Sundrum models, proposed solutions to the hierarchy
problem with one extra dimension. An important distinction between
the radion and the Higgs boson is that the radion would couple directly
to gluon pairs, and in particular its decay products would include
a significant fraction of gluon jets. The radion has the same quantum
numbers as the Standard Model (SM) Higgs boson, and therefore they
can mix, with the resulting mass eigenstates having properties different
from those of the SM Higgs boson. Existing searches for the Higgs
bosons are sensitive to the possible production and decay of radions
and Higgs bosons in these models. For the first time, searches for
the SM Higgs boson and flavour-independent and decay-mode independent
searches for a neutral Higgs boson are used in combination to explore
the parameter space of the Randall-Sundrum model. In the dataset recorded
by the OPAL experiment at LEP, no evidence for radion or Higgs particle
production was observed in any of those searches. The results are
used to set limits on the radion and Higgs boson masses. 

\vfill

{\large 

\bigskip\bigskip\bigskip\bigskip \bigskip\bigskip }{\large \par}

\begin{center}{\large To be submitted to Physics Letters B}\end{center}{\large \par}

\vfill

\end{titlepage}
%
%
\begin{center}{\Large        The OPAL Collaboration
}\end{center}\bigskip
\begin{center}{
G.\thinspace Abbiendi$^{  2}$,
C.\thinspace Ainsley$^{  5}$,
P.F.\thinspace {\AA}kesson$^{  3,  y}$,
G.\thinspace Alexander$^{ 22}$,
J.\thinspace Allison$^{ 16}$,
P.\thinspace Amaral$^{  9}$, 
G.\thinspace Anagnostou$^{  1}$,
K.J.\thinspace Anderson$^{  9}$,
S.\thinspace Asai$^{ 23}$,
D.\thinspace Axen$^{ 27}$
% ...
}\end{center}\bigskip
\bigskip
$^{  1}$School of Physics and Astronomy, University of Birmingham,
Birmingham B15 2TT, UK
\newline
$^{  2}$Dipartimento di Fisica dell' Universit\`a di Bologna and INFN,
I-40126 Bologna, Italy
\newline
$^{  3}$Physikalisches Institut, Universit\"at Bonn,
D-53115 Bonn, Germany
\newline
% ...
$^{  5}$Cavendish Laboratory, Cambridge CB3 0HE, UK
\newline
% ...
$^{  9}$Enrico Fermi Institute and Department of Physics,
University of Chicago, Chicago IL 60637, USA
\newline
% ...
$^{ 16}$Department of Physics, Schuster Laboratory, The University,
Manchester M13 9PL, UK
\newline
% ...
$^{ 22}$Department of Physics and Astronomy, Tel Aviv University,
Tel Aviv 69978, Israel
\newline
$^{ 23}$International Centre for Elementary Particle Physics and
Department of Physics, University of Tokyo, Tokyo 113-0033, and
Kobe University, Kobe 657-8501, Japan
\newline
% ...
$^{ 27}$University of British Columbia, Department of Physics,
Vancouver BC V6T 1Z1, Canada
\newline
% ...
$^{  y}$ now at CERN
\newline
% ...

\pagebreak

\floatplacement{figure}{p}

\bigskip

\section{Introduction}

In \cite{ADDOne}, a model was proposed to solve the problem of the
hierarchy between the electroweak mass scale, $\Lambda_{\mathrm{W}}=\mathcal{O}(\TeV)$,
and the Planck mass $\MPlanck=\mathcal{O}(10^{15}\TeV)$ at which
gravity becomes strong. In this model, the hierarchy is generated
by extending four-dimensional space time with compact extra dimensions.
In the resulting effective four-dimensional theory, $\MPlanck$ appears
enlarged with respect to the hypothesised fundamental value $\widetilde{M}_{\mathrm{Pl}}$,
due to the hidden volume $V_{n}$ of the $n$ extra dimensions:
$\MPlanck^{2}=\widetilde{M}_{\mathrm{Pl}}^{2+n}V_{n}$.
To generate the observed value $\MPlanck=\PlanckScale$ from a hypothesised
fundamental value close to the electroweak scale, $\widetilde{M}_{\mathrm{Pl}}\simeq1\TeV$,
many additional dimensions are necessary or each additional dimension
must be extraordinarily large, which generally conflicts with constraints
from electroweak precision measurements. The constraints do not directly
apply if the electroweak and strong forces and the particles of the
Standard Model (SM) are confined to a four-dimensional subspace (brane),
and only gravity is allowed to propagate into the whole space. Measurements
of the gravitational force limit the size of extra dimensions to $200\mum$
\cite{ExtraDimensionsLimits3}. Model dependent constraints can be
obtained from electroweak precision observables, which can be affected
in a sizable way by gravity \cite{LSGLimits}.

In the Randall-Sundrum (RS) model \cite{RS1,*RS2}, one compact extra
dimension is introduced. As in previous models, the extra dimension
is hidden to the forces and particles of the SM by confining them
to one brane, the SM brane. Only gravity is allowed to propagate into
the extra dimension. In this model the hierarchy is not generated
by the extra volume, but by a specifically chosen {}``warped'' geometry.
As a direct consequence of the geometry, gravity is mainly located
close to a second brane, the Planck brane, which is located at a distance
$r_{0}$ away from the SM brane, and its propagation in the extra
dimension is exponentially damped. Thus, there is only a small overlap
between gravity and SM particles and forces, explaining the weakness
of gravity with respect to the electroweak interaction, i.e.~the
observed mass hierarchy. The constraints on the size of the extra
dimensions do not apply in this case, because the gravitational force
is only weakly modified due to the localisation of gravity.

The model is considered to be a low-energy approximation of a more
fundamental theory and does not explain the mechanism that traps the
SM fields on the brane or the mechanism which gives rise to the geometry.
It is possible to derive models with such a geometry from M-theory
\cite{RSMTheory}. 

The spectrum of the additional particles in the RS model has been
investigated in \cite{RSEinsteinEquation} and \cite{RSRadionDynamics}.
There are massless and massive spin-two excitations. The massless
excitations couple with gravitational strength and can be identified
with gravitons. The masses and couplings of the massive spin-two excitations
are set by the weak scale. These states have not been observed, but
if they exist, they should be observable at experiments using the
next generation of colliders. In addition, there is a spinless excitation,
called the \emph{radion}. The radion corresponds to a local fluctuation
of the inter-brane distance: $r_{0}\ra r_{0}+\Delta r(x)$. To prevent
the branes from drifting apart faster than allowed by cosmological
models, a stabilisation mechanism is needed \cite{RadionStabilisation,*RadionMass,*RadionStabilisation2}.
As a consequence, the radion acquires a mass \cite{RSRadionDynamics}.
To introduce no further hierarchies, the mass should be well below
$1\TeV$. 

The radion carries the same quantum numbers as the Higgs boson; thus
the radion and the Higgs boson can mix. This possibility was investigated
first in \cite{RSEinsteinEquation} and was pursued in \cite{RSRadionDynamics},
where calculations are carried out to higher order. The present study
is based on the Lagrangian of \cite{RSRadionDynamics}. The physical
scalars of the model are derived therein. The couplings to matter
are investigated in \cite{RSEinsteinEquation}, where the calculations
are based on a Lagrangian of a lower order approximation. The ideas
of \cite{RSEinsteinEquation} are transferred to the Lagrangian of
\cite{RSRadionDynamics} leading to the results summarised in Section~\ref{sec:TheRadionPlusTheHiggs}.
The derivation of the physical scalars and the couplings to matter
are detailed in the Appendices~\ref{sec:RSMassEigenstates} and~\ref{sec:RSCouplingToSM}.

Like the SM Higgs boson, both scalars are mainly produced in the {}``Higgsstrahlung''
process, $\ee\ra\Zzero\RSRadion\textrm{ or }\Zzero\RSHiggs$, at LEP2,
where $\RSRadion$ and $\RSHiggs$ are the two scalar mass eigenstates
of the model. The limits on the cross-section of the Higgsstrahlung
process obtained from searches for the SM Higgs boson, flavour independent
searches for hadronically decaying Higgs bosons and decay-mode independent
searches for Higgs bosons are used to restrict the parameter space
of the Randall-Sundrum model as explained in Section~\ref{sec:RSInterpretaion}.

\section{The Scalars of the Randall-Sundrum Model\label{sec:TheRadionPlusTheHiggs}}

In the Randall-Sundrum model there are two scalar particles, the radion
and the Higgs boson. Their masses, $\RSmr$ and $\RSmh$, are free
parameters. Further free parameters are: $\RSScale$, which sets the
mass scale on the SM brane and is expected to be $\EuScript{O}(1\TeV)$,
and $\xi$ which controls the kinematic mixing between the radion
and the Higgs boson. 

The radion couples to the trace of the energy momentum tensor. Thus,
to first order the radion couples to massive particles with couplings
proportional to the particle mass, and the Lorentz structure of the
couplings is identical to that of the Higgs boson. However, the coupling
strength of the radion is generally reduced by $\VEV/\sqrt{6}\RSScale$
w.r.t. the couplings of the SM Higgs boson, where $\VEV$ denotes
the vacuum expectation value of the Higgs field. Unlike the Higgs
boson, which only couples to gluons via a top loop, the radion couples
directly to gluon pairs due to the anomaly of the trace of the energy
momentum tensor. As a consequence, the radion decays mostly into gluon
pairs. 

\begin{figure}[p]
  \begin{center}
    \includegraphics{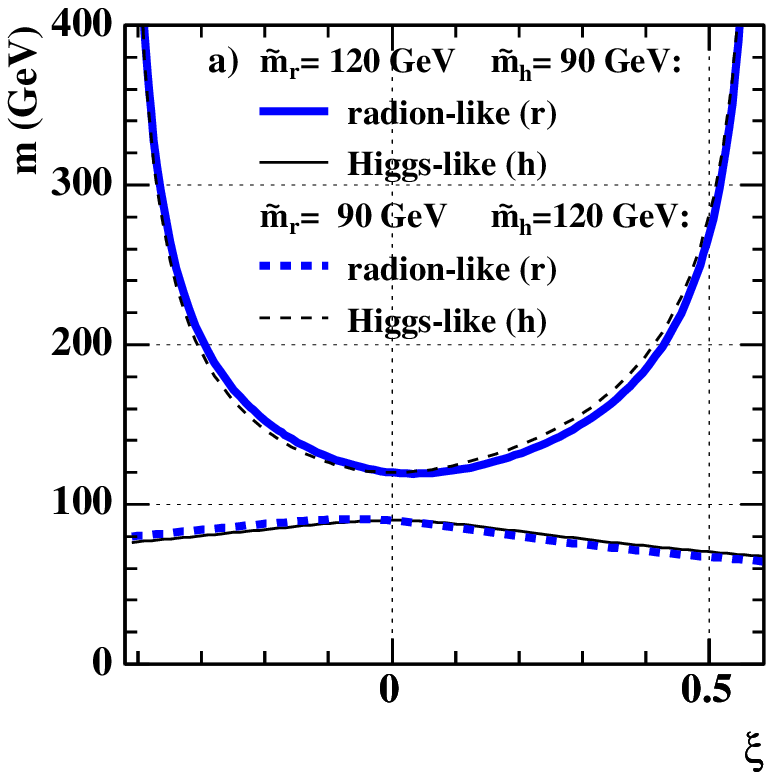}
    \hfill
    \includegraphics{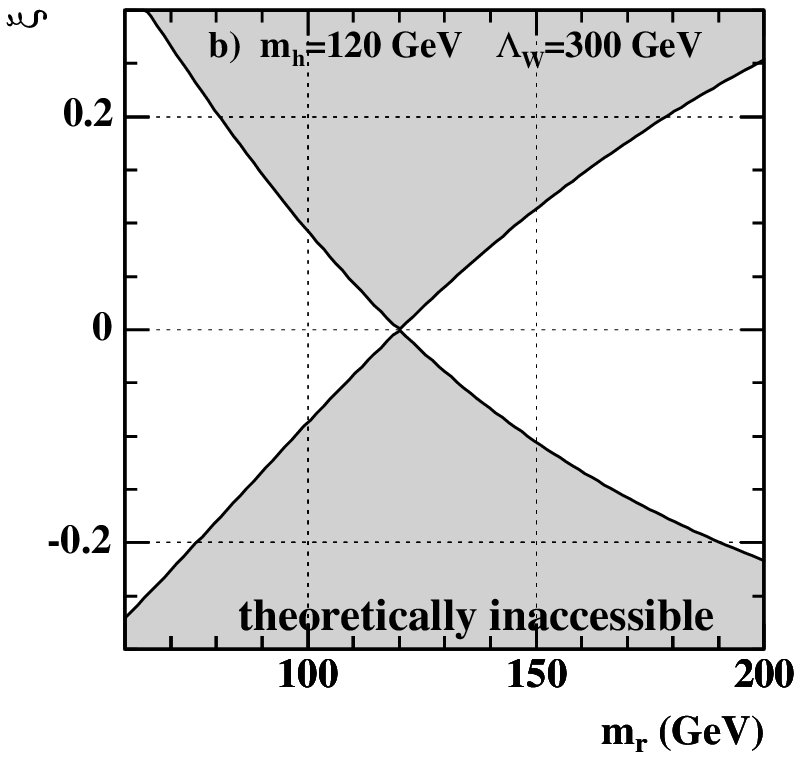}
  \end{center}

  \caption{\label{fig:RadionEigenMasses}a) Masses $m_{\mathrm{r}/\mathrm{h}}$
     of the heavy and light mass eigenstates for fundamental Higgs boson
     and radion mass parameters, $\RSFundmh$ and $\RSFundmr$, of $90\GeV$
     and $120\GeV$. The fundamental radion is chosen to be heavier (lighter)
     than the Higgs boson, indicated by the solid (dashed) lines. The $x$-axis
     extends over the allowed $\xi$-range. b) Allowed parameter space
     in the $m_{\mathrm{r}}$ and $\xi$ plane for a Higgs boson mass $\RSmh=120\GeV$.
     Outside the permitted region the Higgs and radion-like states are
     unphysical (ghost-like). In both figures the weak scale was chosen
     to be $\Lambda_{\mathrm{W}}=300\GeV$.}
\end{figure}

Due to the kinematic mixing of the radion and the Higgs boson, both
physical scalars, the Higgs-like and the radion-like state $\RSHiggs$
and $\RSRadion$, may have properties different from those of the
SM Higgs boson. Here, the radion-like and the Higgs-like states, $\RSRadion(\xi)$
and $\RSHiggs(\xi)$, are defined such that the Higgs-like state becomes
the SM Higgs boson in the limit $\xi\ra0$, and the mapping between
the fundamental mass parameters (the mass parameter of the Higgs mechanism,
$\RSFundmh$, and the mass parameter assigned to the radion excitation,
$\RSFundmr$) to the mass eigenvalues is a continuous function of
$\xi$ (see Figure~\ref{fig:RadionEigenMasses}a and Appendix~\ref{sec:RSMassEigenstates}
for details). 

For non-zero mixing ($\xi\neq0)$ some combinations of the masses
$\RSmr$ and $\RSmh$ of the radion-like and the Higgs-like state
will lead to unphysical particles (ghosts or tachyons). The allowed
minimum and maximum mixing is limited by requiring the particles to
be physical. The limits depend on the masses, $\RSmr$ and $\RSmh$,
and the mass scale $\RSScale.$ For fixed masses, the bounds increase
with $\RSScale$. The physical regions are displayed in Figure~\ref{fig:RadionEigenMasses}b
as a function of the mixing parameter $\xi$, and $\RSmr$ for one
$\RSScale$ and $\RSmh$.

Both particles, the radion and the Higgs boson, are predominantly
produced in {}``Higgsstrahlung'' in $\ee$ collisions for masses
in the range accessible by the LEP experiments. The production of
the radion-like and the Higgs-like states are complementary as seen
in Figure~\ref{fig:RadionCrossSection}a and~b. The branching ratio
of the Higgs-like state into heavy quarks and leptons may be reduced
depending on the mixing parameter $\xi$ while the branching ratio
into gluon pairs is enhanced, which can be seen in Figure~\ref{fig:RadionCrossSection}c
and~d. Therefore, searches for the SM Higgs boson (assuming $\mHsm\ll2\mW$)
which are sensitive only to the decay mode $\mathrm{h}\ra\bb$, may
lose their sensitivity, in contrast to flavour independent searches
which are sensitive to $\mathrm{h\ra\glgl}$.

\section{Experimental Constraints on the Randall-Sundrum Model\label{sec:RSInterpretaion}}

\begin{figure}[p]
  \begin{center}
     \includegraphics{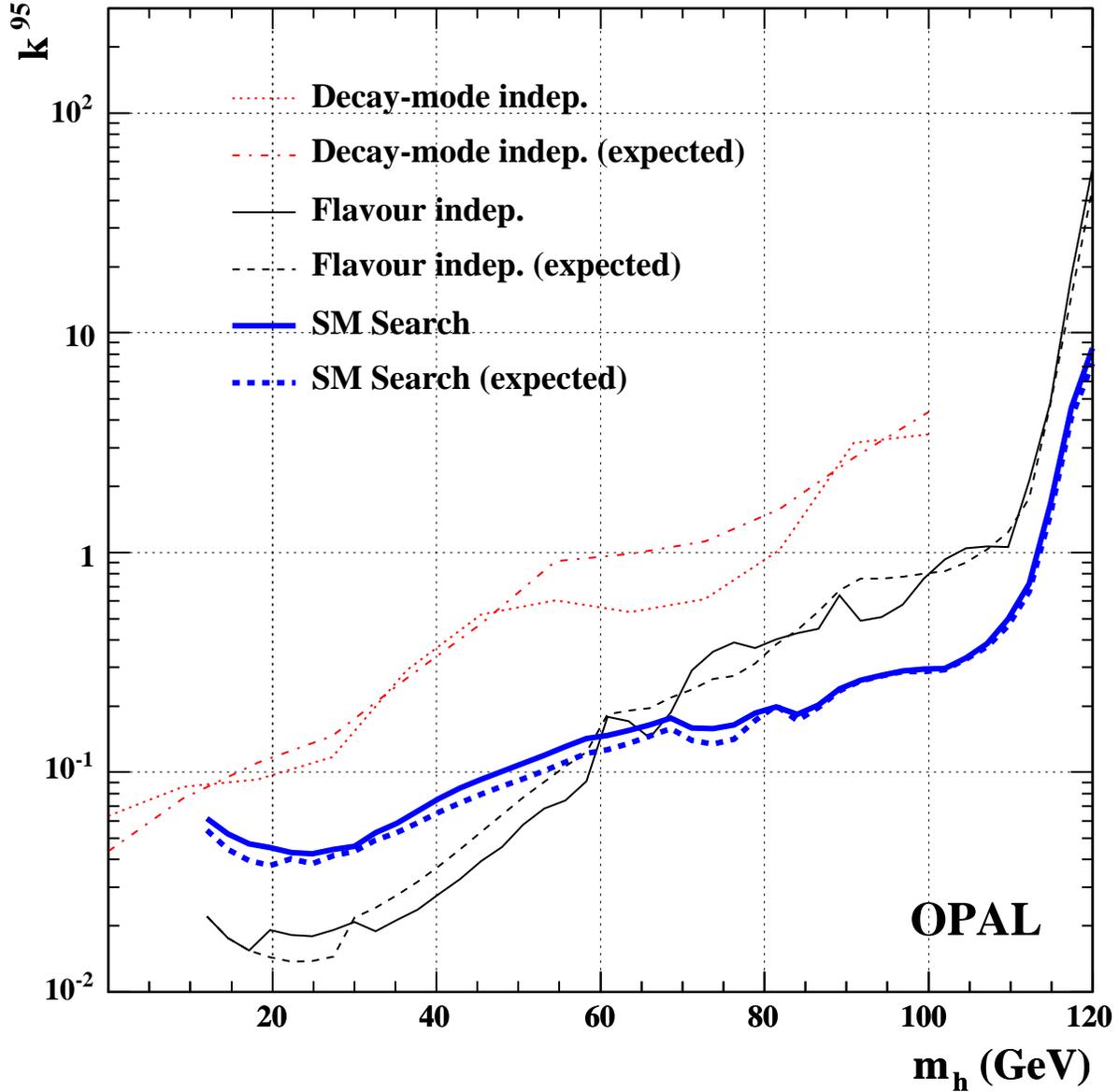}
  \end{center}

  \caption{\label{fig:Limits}The observed and expected limits on the scale
     factor k as a function of the Higgs boson mass obtained by the SM
     Higgs boson search, the flavour independent and the decay-mode independent
     Higgs boson search. The scale factor k relates the cross-section times
     branching ratio to the cross-section of SM Higgsstrahlung. The limits
     equally apply to the radion-like and the Higgs-like state of the Randall-Sundrum
     model each with the mass $\RSmh$.}
\end{figure}

Since the signatures of the radion-like and the Higgs-like states
are similar to the signatures of the SM Higgs boson or neutral Higgs
bosons of more general models, searches for a neutral Higgs boson
also constrain the parameter space of the Randall-Sundrum (RS) model.
The following searches for the Higgsstrahlung process, $\ee\ra\Zzero\mathrm{\varphi}$,
are exploited, where $\varphi$ is a scalar:

\begin{enumerate}
\item The search for the SM Higgs boson \cite{opalSMHiggsFinal}, $\varphi=\Hsm$,
which exploits the properties of the dominant decay mode of the SM
Higgs boson, $\mathrm{\Hsm\ra\bb}$ (assuming $\mHsm\ll2\times\mW$).
The decay $\Hsm\ra\tautau$ is not considered here. The search uses
$593\invpb$ and $170\invpb$ of data collected with the OPAL detector
at $\sqrts=189-209\GeV$ and $\sqrts=91\GeV$, respectively. All possible
decay modes of the $\Zzero$ boson are considered: 
$\Zzero\ra\qq,\,\ee,\,\mu^{+}\mu^{-},\,\tautau\textrm{ and }\nu\bar{\nu}$.
\item A flavour independent search for hadronically decaying Higgs bosons,
$\varphi=\hzero$, sensitive to the $\RSHiggs\ra\qq$ and $\RSHiggs\ra\glgl$
modes, using the same dataset as above \cite{opalFlavIndep}.
\item A search \cite{OPALHiggsDecayModeIndep}, independent of the decay
mode of the scalar particle, using events in which the $\Zzero$ boson
decays into muon or electron pairs. There are no assumptions on the
scalar particle decay. Although this search gives weaker limits than
the two above, it is the only search to cover the mass region from
$1\MeV$ to $12\GeV$.
\end{enumerate}
These searches have not revealed any significant excess of data over
the background from Standard Model processes, and limits on the cross-section
of the Higgsstrahlung process times the branching ratio of the scalar
particle decay have been derived at the $95\%$ confidence level.
The limits are expressed in terms of a scaling factor $k_{\varphi x}^{95}$,
which relates the maximally allowed cross-section times branching
ratio, $\sigma_{\varphi\Zzero}^{95}(m_{\varphi})\times\mathrm{Br}(\varphi\ra x\bar{x})$,
of a scalar particle $\varphi$ to the expectation for Higgs boson
production$\sigma_{\mathrm{HZ}}^{\mathrm{SM}}(m_{\varphi})$ from
the SM:
\begin{equation}
k_{\varphi x}^{95}(m_{\varphi})\deq
  \frac{\sigma_{\mathrm{\varphi Z}}^{95}(m_{\varphi})}
       {\sigma_{\mathrm{HZ}}^{\mathrm{SM}}(m_{\varphi})}\times\mathrm{Br}
  (\varphi\ra x\bar{x})
\label{eq:kFractor}
\end{equation}
A value $k_{\varphi x}^{95}(m_{\varphi x})=1$ means that at the $95\%$
confidence level, a cross-section could be excluded which is equal
to the cross-section of the Higgsstrahlung process, $\ee\ra\Hsm\Zzero$,
for a SM Higgs boson $\Hsm$ having the mass $m_{\varphi}$. The observed
and expected limits are depicted in Figure~\ref{fig:Limits}. The
first search is sensitive only to $\varphi\ra\bb$, the second to
$\varphi\ra\qq$, $\varphi\ra\glgl$, and the third analysis covers
all possible decays. 

In the RS model, the radion-like and the Higgs-like states have the
same coupling structure as a SM Higgs boson. The couplings to fermions
$f$ or vector bosons $V$ only differ by factors $\sqrt{k_{f}}$
or $\sqrt{k_{V}}$ which depend on the masses of the radion-like and
the Higgs-like states, $\RSmr$ and $\RSmh$, the mixing parameter
$\xi$, and the mass scale $\RSScale$ (see Appendix~\ref{sec:RSCouplingToSM}).
Thus, the limits $k_{\varphi x}^{95}$ apply to the processes predicted
in the RS model, $\ee\ra\Zzero\varphi$, where $\varphi$ is the radion-like
state $\RSRadion$ or the Higgs-like state $\RSHiggs$. 

\begin{figure}[p]
  \begin{minipage}{\textwidth/2}
     \centering{\bf \hspace{1cm}SM Higgs Search Limits}

     \includegraphics[width=8cm]{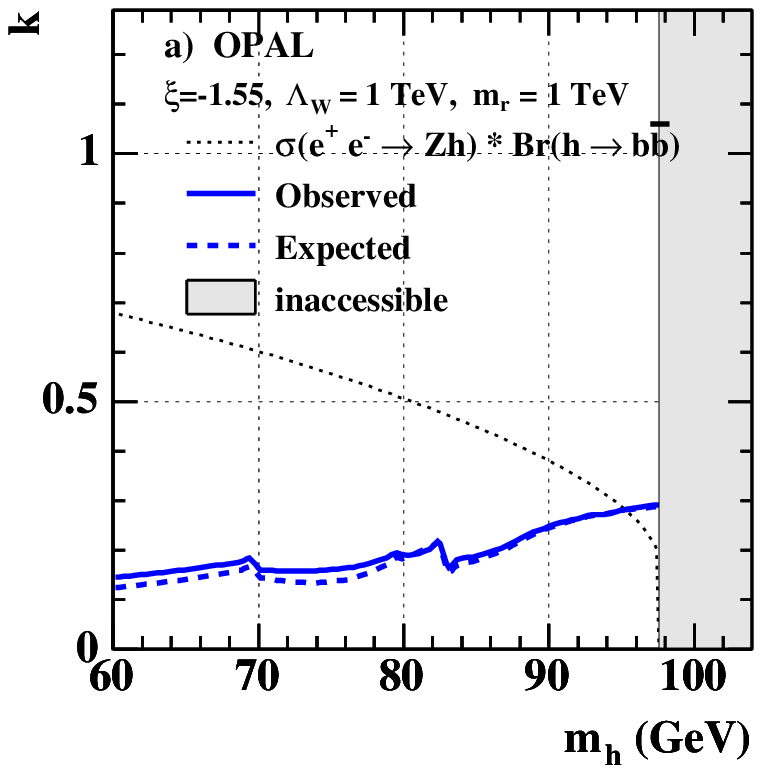}
  \end{minipage}
  \hfill
  \begin{minipage}{\textwidth/2}
     \centering{\bf \hspace{1cm}Flavour Independent Limits}
     \includegraphics[width=8cm]{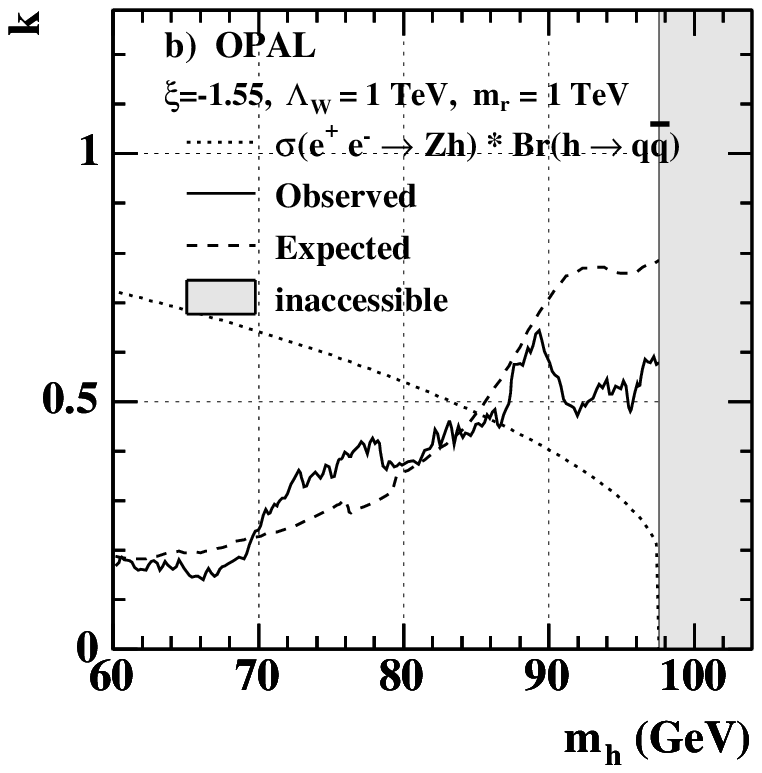}
  \end{minipage}

  \vspace{.3cm}

  \begin{minipage}{\textwidth/2}
     \centering{\bf \hspace{1cm}SM Higgs Search Limits}
     \includegraphics[width=8cm]{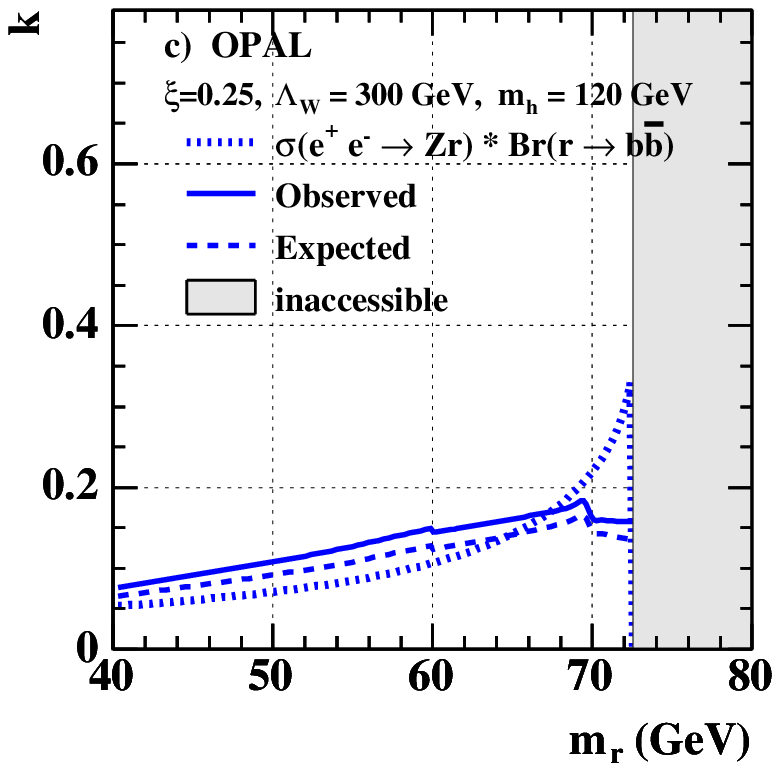}
  \end{minipage}
  \hfill
  \begin{minipage}{\textwidth/2}
     \centering{\bf \hspace{1cm}Flavour Independent Limits}
     \includegraphics[width=8cm]{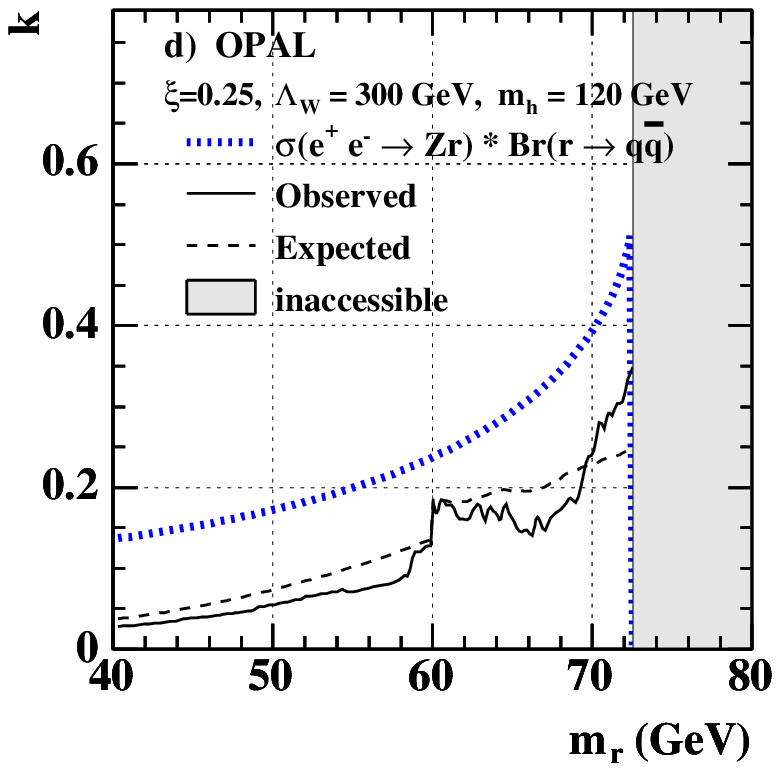}
  \end{minipage}

  \caption{\label{fig:RSLimitAndPrediction}The cross-section times branching
     ratio of the Higgs-like (Figure~a and~b) and radion-like state (Figure~c
     and~d) relative to the cross-section of SM Higgsstrahlung together
     with the observed and expected limits (solid and dashed lines) obtained
     from the SM (Figure~a and~c) and the flavour independent (Figure~b
     and~d) Higgs boson searches at one point in the Randall-Sundrum parameter
     space as a function of the mass of the Higgs-like state $\RSmh$ and
     the mass of the radion-like state $\RSmr$. The dotted lines in Figures~a
     and~c indicate the cross-section times
     $\mathrm{Br}(\RSRadion\textrm{ or }\RSHiggs\ra\bb)$
     and in Figures~b and~d the cross-section times 
     $\mathrm{Br}(\RSRadion\textrm{ or }\RSHiggs\ra\mathrm{hadrons})$.
     The shaded region is inaccessible by the theory. Model points are
     excluded if the predicted cross-section times branching ratio exceeds
     the limit.}
\end{figure}

Points in the parameter space of the RS model are considered excluded
if the predicted cross-section times branching ratio for either the
radion-like or the Higgs-like state exceeds the limit obtained from
one of the Higgs boson searches. At each scan point, the search is
chosen which yields the most restrictive expected limit. For example
in Figures~\ref{fig:RSLimitAndPrediction}a-d, the cross-sections
times branching ratio of the radion-like and Higgs-like state are
shown together with the limit obtained from the flavour independent
and the SM Higgs boson search. For the model points of Figures~\ref{fig:RSLimitAndPrediction}a
and b, a small region in the parameter space just before the inaccessible
region remains allowed. Neither the SM nor the flavour independent
Higgs boson search is able to exclude this region. For the parameters
shown in Figure~\ref{fig:RSLimitAndPrediction}c, the SM search is
not capable of excluding the model points for the parameters $\xi=0.25$,
$\RSScale=300\GeV$, $\RSmh=120\GeV$, and for masses of the radion-like
state $\RSmr\lesssim67\GeV$. The flavour independent Higgs boson
search excludes all model points up to the inaccessible region (Figure
~\ref{fig:RSLimitAndPrediction}d). 

\begin{figure}[p]
  \begin{center}
     \includegraphics[
                      width=16cm,
                      keepaspectratio]%
                     {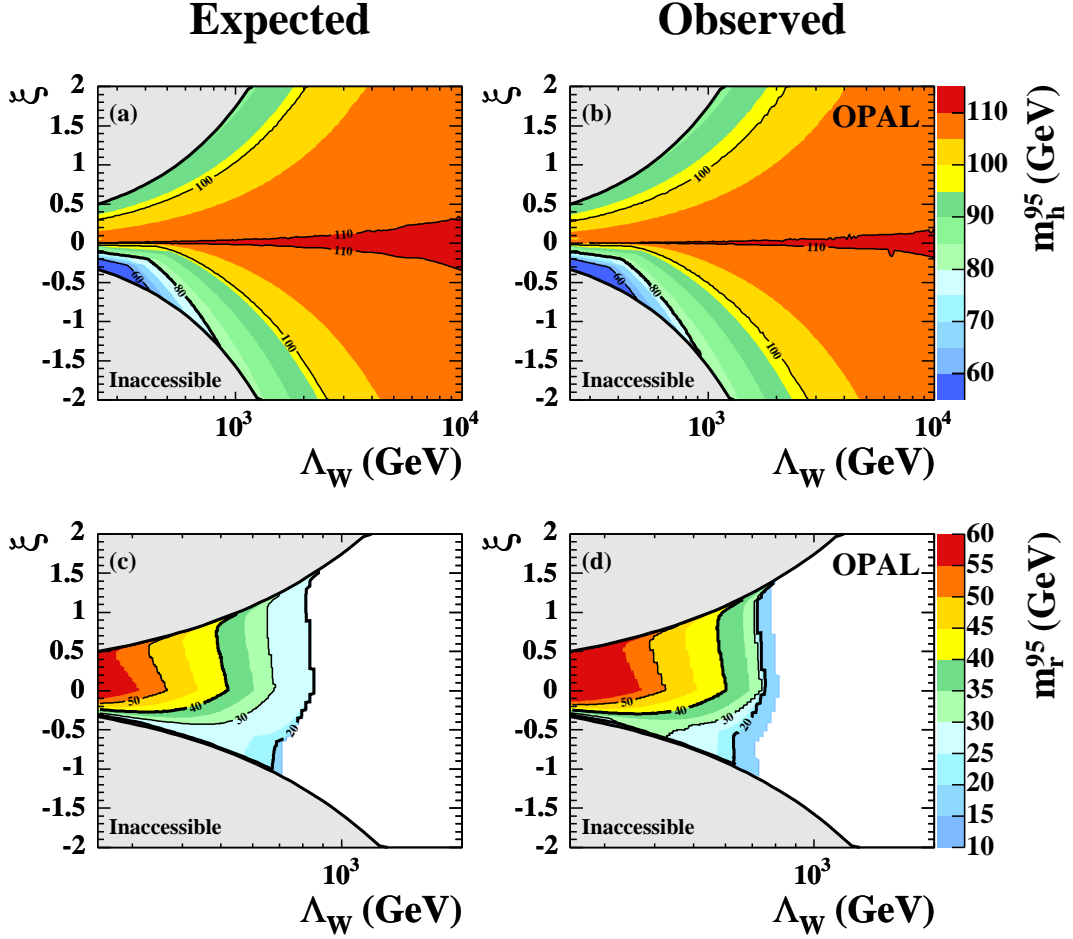}
  \end{center}

  \caption{\label{fig:RSMassLimitPlanes}Expected and observed lower limits
     on the mass of the Higgs-like and the radion-like state, $\RSmh$
     (a and~b) and $\RSmr$ (c and~d), as a function of the mixing parameter
     $\xi$ and the scale parameter $\RSScale$. The Figures~a) and~c)
     show the expected limit, and Figures~b) and~d) the observed limit.
     Inside each shaded region, the obtained lower mass limit is equal
     or larger than the value indicated by the code on the right. The regions
     in the upper and lower left corner are inaccessible by the theory. }
\end{figure}

To find the lowest masses compatible with the observations, scans
over the parameter space of the RS model are performed. Figures~\ref{fig:RSMassLimitPlanes}a
and~b show the lowest mass of the Higgs-like state allowed at the
$95\%$ confidence level in the plane spanned by the mixing parameter
$\xi$ and the scale parameter $\RSScale$. In the $\xi$-direction
an equidistant grid is chosen using $200$ points between the minimum
and maximum value of the allowed region. In the $\RSScale$-direction,
$160$ scan points are chosen equally spaced on a logarithmic scale
from $246\GeV$ to $10\TeV$. At each scan point, $\RSmr$ is scanned
initially in coarse steps in the range from $1\MeV$ to $1\TeV$,
where the step sizes are $1-\RSMScanStepLow$ and $\RSMScanStepHigh$
below and above $400\GeV$, respectively. For each $\RSmr$ value,
$\RSmh$ is scanned in the range from $1\MeV$ to $120\GeV$ in steps
of $1\GeV$. The scan stops if the predicted cross-section times branching
ratio of both the radion-like and the Higgs-like states drops below
the limit of the most sensitive Higgs boson search. Finally, the mass
$\RSmh$ at which the cross-section drops below the limit is found
to within $250\MeV$ by an iterative procedure.

\begin{figure}[p]
  \begin{center}
     \includegraphics[width=12cm,
                      keepaspectratio]%
                     {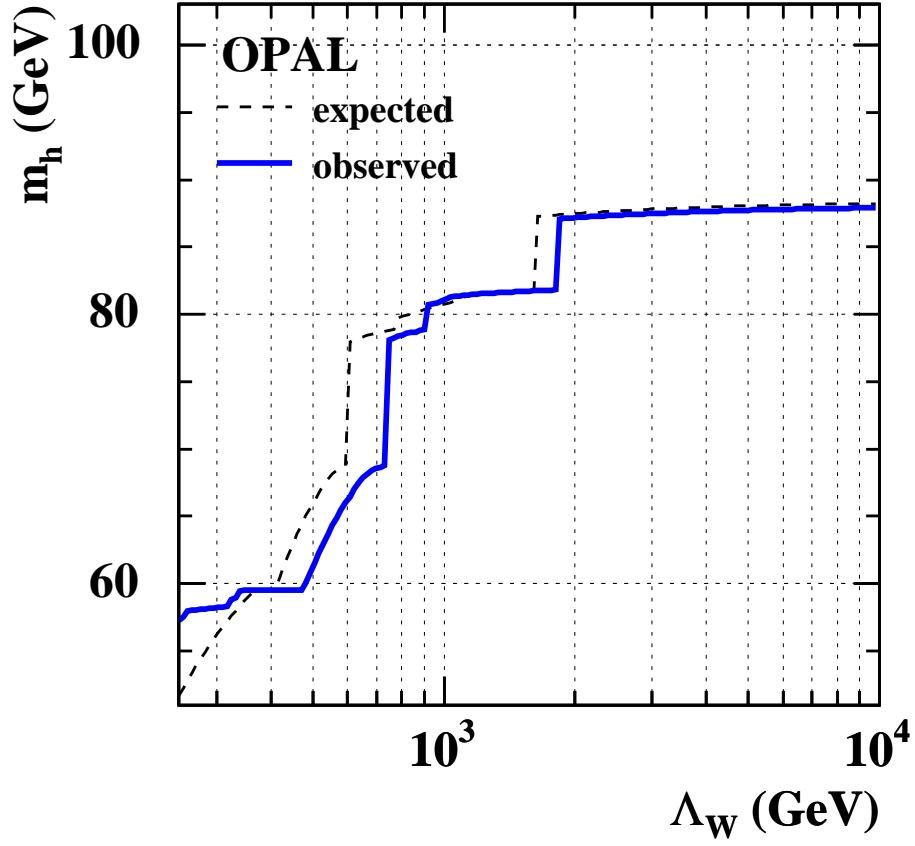}
  \end{center}

  \caption{\label{fig:RSUpperMhLimit}The lowest expected and observed limit
     on the Higgs boson mass as a function of the scale parameter $\RSScale$
     for all allowed $\xi$ and for masses of the radion-like state $\RSmr$
     in the range from $1\MeV$ to $1\TeV$. The analyses often lose their
     sensitivity close to the inaccessible region. If the region up to
     the inaccessible region is covered, the next allowed mass will be
     several $\GeV$ further away. This causes the step like structure.}
\end{figure}

For zero mixing ($\xi=0$), the mass limit of the SM Higgs boson search
is obtained. For non-zero mixing, the mass limit of the Higgs-like
state is generally lower and decreasing with decreasing scale parameter
$\RSScale$. The lowest mass limits are generally obtained for maximum
or minimum values of $\xi$ and values of the radion mass much larger
than the limit on $\RSmh$. In Figure~\ref{fig:RSUpperMhLimit} the
lowest mass limits of the Higgs-like states are shown for all $\xi$
allowed by the theory. At large $\RSScale$, the maximally allowed
$|\xi|$ is beyond $\mathcal{O}(1)$. For all $\xi$, $\RSmr$ and
$\RSScale$, the Higgs mass has to be larger than $\RSMhLimitMinObs$
at the $95\%$ confidence level, where a limit of $\RSMhLimitMinExp$
is expected. In cases in which either the observed limit or the expected
limit is obtained just before the inaccessible region, the difference
between the observed and expected limit may become large, if one of
them is beyond and the other just before the inaccessible region.
If for example in Figure~\ref{fig:RSLimitAndPrediction}b, the cross-section
was slightly higher such that it was just above the observed cross-section
limit and it crossed the expected limit at $90\GeV$, the expected
limit on $\RSmh$ would have been at $90\GeV$ and the observed limit
would have been beyond the inaccessible region which would yield a
limit larger than $100\GeV$. This leads to the large steps in Figure~
\ref{fig:RSUpperMhLimit}. 

The same procedure was performed to find the lowest allowed mass of
the radion-like state, $\RSmr$. The result of the scan in the $\xi-\RSScale$
plane is shown in Figures~\ref{fig:RSMassLimitPlanes}c and~d. The
cross-section of the radion-like state vanishes for large negative
mixing and decreases rapidly with increasing $\RSScale$, since the
couplings of the radion to SM particles is proportional to the inverse
of $\RSScale$. The analyses lose their sensitivity for $\RSScale\gtrsim\RSMrSensitivLambdaMax$
and for maximal negative mixing; therefore, a mass limit independent
of the mixing parameter $\xi$ cannot be extracted.

\section{Summary}

Limits on the Higgsstrahlung cross-section obtained from data recorded
with the OPAL detector have been used to restrict the parameter space
of the Randall-Sundrum model. The data exclude masses for the Higgs-like
state below $\RSMhLimitMinObs$ for all scales $\RSScale\geq246\GeV$,
independent of the mixing between the radion and the Higgs boson,
and of the radion mass. The analyses are sensitive to the radion for
scales $\RSScale\lesssim\RSMrSensitivLambdaMax$. No universal limit,
independent of $\RSScale,$ $\xi$ and $\RSmh$, on the mass of the
radion-like state can be extracted.

\section*{Acknowledgements}

We particularly wish to thank Laura Covi, Csaba Cs\'aki, Graham D. Kribs, 
Thomas G. Rizzo and James Wells for their support. 
We also thank the SL Division for the efficient operation
of the LEP accelerator at all energies
 and for their close cooperation with
our experimental group.  In addition to the support staff at our own
institutions we are pleased to acknowledge the  \\
Department of Energy, USA, \\
National Science Foundation, USA, \\
Particle Physics and Astronomy Research Council, UK, \\
Natural Sciences and Engineering Research Council, Canada, \\
Israel Science Foundation, administered by the Israel
Academy of Science and Humanities, \\
Benoziyo Center for High Energy Physics,\\
Japanese Ministry of Education, Culture, Sports, Science and
Technology (MEXT) and a grant under the MEXT International
Science Research Program,\\
Japanese Society for the Promotion of Science (JSPS),\\
German Israeli Bi-national Science Foundation (GIF), \\
Bundesministerium f\"ur Bildung und Forschung, Germany, \\
National Research Council of Canada, \\
Hungarian Foundation for Scientific Research, OTKA T-038240, 
and T-042864,\\
The NWO/NATO Fund for Scientific Research, the Netherlands.\\

\appendix

\section{Appendix}

\renewcommand{\arraystretch}{1.3}

\subsection{Physical Scalars in the RS-Model\label{sec:RSMassEigenstates}}

In \cite{RSRadionDynamics}, the effective 4D Lagrangian is derived,
which describes the kinetic terms of the radion and the Higgs boson
and their couplings to SM particles. Starting from the effective Lagrangian,
the physical states and their masses are computed as shown in \cite{RSRadionDynamics},
and the radion-like and Higgs-like states are defined.

The following kinetic terms for the radion $\RSFundr$ and the Higgs
boson $\RSFundh$ have been found:
\begin{equation}
  \mathcal{L}_{\mathrm{scalar}}\simeq\left(\begin{array}{c}
     \RSFundh\\
     \RSFundr
  \end{array}\right)^{\mathrm{T}}\left(\begin{array}{cc}
     -\frac{1}{2}\square-\frac{1}{2}\RSFundmh^{2} & 
     3\xi\gamma\square\\
     3\xi\gamma\square & 
     -\frac{1}{2}\left(1+6\xi\gamma^{2}\right)\square-\frac{1}{2}\RSFundmr^{2}
  \end{array}\right)\left(\begin{array}{c}
     \RSFundh\\
     \RSFundr
  \end{array}\right),
\label{eq:RSHiggsRadionLagrange}
\end{equation}
 where $\xi$ is a free parameter of $\mathcal{O}(1)$, leading to
the kinetic mixing between the radion and the Higgs boson. The normalisation
of the radion field depends on $\gamma\deq\VEV/\sqrt{6}\RSScale$,
where $\VEV$ is the vacuum expectation value of the Higgs field and
$\RSScale$ the mass scale on the SM brane. The values $\RSFundmr$
and $\RSFundmh$ are fundamental mass parameters of the radion and
the Higgs fields. 

The physical states are obtained by diagonalisation of the matrix
in Equation~(\ref{eq:RSHiggsRadionLagrange}) \cite{RSRadionDynamics}.
First the kinetic mixing is resolved by the choice $\RSFundh\deq h^{\prime}+6\xi\gamma r^{\prime}/Z$
and $\RSFundr\deq r^{\prime}/Z$, with:
\begin{equation}
  Z\deq\sqrt{1+6\xi\gamma^{2}(1-6\xi)}.
  \label{eq:RSDefZ}
\end{equation}
The fields, $h^{\prime}$ and $r^{\prime}$, are real i.e.~physical
scalars only if:
\begin{equation}
  \frac{1}{12}\left(1-\sqrt{1+\frac{4}{\gamma^{2}}}\right)
  <\xi
  <\frac{1}{12}\left(1+\sqrt{1+\frac{4}{\gamma^{2}}}\right).
  \label{eq:RSXiCondition}
\end{equation}
 The choice of $h^{\prime}$ and $r^{\prime}$ removes the kinetic
mixing, but introduces a mixing of the mass terms for non zero $\RSFundmr$
and $\RSFundmh$. The matrix of the mass terms is diagonalised by
rotating by the angle $\theta$:
\begin{equation}
  \tan2\theta\deq12\xi\gamma Z\frac{\RSFundmh^{2}}{\RSFundmr^{2}-\RSFundmh^{2}(Z^{2}-36\xi^{2}\gamma^{2})}.
  \label{eq:RSRotationAngle}
\end{equation}
The canonically normalised kinetic terms of the fields $h^{\prime}$
and $r^{\prime}$ are invariant under rotations. The full transformation
yields the following relations between the fundamental states, $\RSFundh$
and $\RSFundr$, and the mass eigenstates, $\RSEigenh$ and $\RSEigenr$:
\begin{eqnarray}
  \RSFundh & 
    = & 
    (\cos\theta-\frac{6\xi\gamma}{Z}\sin\theta)\RSEigenh+(\sin\theta+\frac{6\xi\gamma}{Z}\cos\theta)\RSEigenr
  \label{eq:RadionMassEigenstates}\\
  \RSFundr & 
    = & 
    -\sin\theta\frac{\RSEigenh}{Z}+\cos\theta\frac{\RSEigenr}{Z}.
  \nonumber 
\end{eqnarray}
The corresponding masses are given by $m_{\pm}$, where $m_{-}\leq m_{+}$
($m_{-}=m_{+}$ for $\xi=0$ and $\RSFundmr=\RSFundmh$):
\begin{equation}
  m_{\pm}^{2}\deq
    \frac{1}{2Z^{2}}\left(
       \RSFundmr^{2}+(1+6\xi\gamma^{2})\RSFundmh^{2}
       \pm\sqrt{\left(\RSFundmr^{2}-\RSFundmh^{2}(1+6\xi\gamma^{2})\right)^{2}
	     +144\gamma^{2}\xi^{2}\RSFundmr^{2}\RSFundmh^{2}}
    \right).
  \label{eq:RSMassEigenValues}
\end{equation}
For $\xi=0$, $m_{+}$ is the mass of the mass eigenstate $\RSEigenh$
if $\RSFundmh>\RSFundmr$ (otherwise this is the mass of the eigenstate
$\RSEigenr$). The assignment of $m_{\pm}$ to the eigenstates $\RSEigenr$
and $\RSEigenh$ changes at the poles, $\xi_{0}$, of (\ref{eq:RSRotationAngle}):
$\RSFundmr=\RSFundmh(Z^{2}-36\xi_{0}^{2}\gamma^{2})$. Here, the rotation
angle $\theta$ flips by $\pi/2$. For $|\xi|>|\xi_{0}|$, $\RSEigenh$
becomes eigenstates with mass $m_{-}$ if $\RSFundmh>\RSFundmr$
(otherwise of the eigenstate $\RSEigenr$).

In the following, the radion-like and Higgs-like state, $\RSRadion$
and $\RSHiggs$, are defined such that for $\xi=0$ the fundamental
radion $\RSFundr$ and the mass eigenstate $\RSRadion$ coincide,
and furthermore, the mass $\RSmr$ and the couplings (see Section~\ref{sec:RSCouplingToSM})
are continuous functions of $\xi$. The definition of $\RSRadion$
is:
\begin{equation}
  \RSRadion\deq
     \left\{ \begin{array}{cc}
        \RSEigenr & 
        \quad\begin{array}{c}
            \textrm{if}\,\quad(\,\RSFundmr>\RSFundmh
	       \textrm{ and }\xi^{2}<\frac{\RSFundmh Z^{2}-\RSFundmr}{36\gamma^{2}\RSFundmh}\,)\\
            \,\,\textrm{or}\,\,(\,\RSFundmr\leq\RSFundmh
               \textrm{ and }\xi^{2}\geq\frac{\RSFundmh Z^{2}-\RSFundmr}{36\gamma^{2}\RSFundmh}\,)
        \end{array}\\
        \RSEigenh & 
        \textrm{otherwise}
    \end{array}\right.
    \begin{array}{c}\\.
    \end{array}
  \label{eq:RSRadionDefinition}
\end{equation}
The corresponding mass is $\RSmr=m_{-}$ if $\RSFundmr\leq\RSFundmh$
and $\RSmr=m_{+}$ if $\RSFundmr>\RSFundmh$. The Higgs-like state
and its mass are defined accordingly. The masses are shown in Figure~\ref{fig:RadionEigenMasses}a
as a function of $\xi$ for fundamental radion and Higgs boson mass
parameters $\RSFundmr$ and $\RSFundmh$ of $90\GeV$ and $120\GeV$.

Equations~(\ref{eq:RSMassEigenValues}) can be solved for $\RSFundmr$
and $\RSFundmh$:
\begin{eqnarray}
  \RSFundmr^{2} & 
    = & 
    \frac{Z^{2}}{2}\left(\MrhQ\pm\sqrt{\MrhD}\right)\nonumber \\
  \RSFundmh^{2} &
    = & 
    \frac{Z^{2}}{2\left(1+6\xi\gamma^{2}\right)}\left(\MrhQ\mp\sqrt{\MrhD}\right).
  \label{eq:RSInvertedMassRelation}
\end{eqnarray}
The signs have to be chosen such that $\RSmr(\xi=0)=\RSFundmr$ and
$\RSmh(\xi=0)=\RSFundmh$. The computed masses $\RSFundmr$ and $\RSFundmh$
are real only if: 
\begin{equation}
  \frac{m_{+}^{2}}{m_{-}^{2}}
  \ge
  \frac{1}{Z^{2}}\left(1+6\xi\gamma^{2}(1+6\xi)+12\gamma\sqrt{\xi^{2}(6\xi\gamma^{2}+1)}\right).
  \label{eq:RadionInversionCondition}
\end{equation}
This condition, together with (\ref{eq:RSXiCondition}), limits the
possible physical parameters as illustrated in Figure~\ref{fig:RadionEigenMasses}b.

\subsection{Couplings of the Higgs Boson and Radion to SM Particles\label{sec:RSCouplingToSM}}

The couplings of the radion-like and the Higgs-like states, which
are defined in the Appendix~\ref{sec:RSMassEigenstates}, are extracted
applying the methods of \cite{RSEinsteinEquation}. In contrast to
\cite{RSEinsteinEquation}, the physical states are derived from the
effective Lagrangian of \cite{RSRadionDynamics} , which is a higher
order approximation.

The radion couples to the trace of the energy-momentum tensor $T_{\mu}^{\,\mu}$
\cite{RSEinsteinEquation}; therefore, the couplings to matter are
similar to those of the SM Higgs boson at lowest order since:
\begin{equation}
  T_{\mu}^{\,\mu}
    =
    -(m_{ij}\bar{\psi}_{i}\psi_{j}-m_{\mathrm{V}}\mathrm{V}_{\mu}\mathrm{V}^{\mu})+\ldots\,,
  \label{eq:EMTensorTrace}
\end{equation}
 where $\psi_{i}$ and $\mathrm{V_{\mu}}$ denote fermions and bosons,
$m_{ij}$ and $m_{\mathrm{V}}$ their masses. The contribution of
terms with derivatives of fields or more than two fields is negligible
here. The combined interaction term of the radion and the Higgs boson
is:
\begin{equation}
  \mathcal{L}_{\mathrm{radion/Higgs}\,\mathrm{inter.}}
    \simeq
    -\frac{1}{v}(m_{ij}\bar{\psi}_{i}\psi_{j}-m_{\mathrm{V}}\mathrm{V}_{\mu}\mathrm{V}^{\mu})
     \left[\RSFundh-\gamma\RSFundr\right],
   \label{eq:RadionParticleInteraction}
\end{equation}
where $\VEV$ denotes the vacuum expectation value of the Higgs field.
The couplings of the radion to the fermions and bosons are generally
reduced by the factor $\gamma=\VEV/\sqrt{6}\RSScale$ compared to
the corresponding couplings of the Higgs boson. 

The couplings of the radion-like and the Higgs-like state $\RSRadion$
and $\RSHiggs$ are obtained by inserting (\ref{eq:RadionMassEigenstates})
according to (\ref{eq:RSRadionDefinition}) into (\ref{eq:RadionParticleInteraction})
and comparing the resulting terms with the Higgs interaction terms
of the SM Lagrangian. This yields for the radion-like state, expressed
in terms of the partial decay width relative to the one of the SM
Higgs boson\footnote{%
For a given mass $\RSmr$ ($\RSmh$) the expression has to be evaluated
using a mass $\mHsm=\RSmr$ ($\mHsm=\RSmh$).}:
\begin{equation}
  k_{\mathrm{f}}=k_{\mathrm{V}}
  \deq
   \frac{\Gamma(\RSRadion\ra\bar{\mathrm{f}}\mathrm{f})}
        {\Gamma(\Hsm\rightarrow\bar{\mathrm{f}}\mathrm{f})}
  =\frac{\Gamma(\RSRadion\textrm{ }\ra\mathrm{VV})}
        {\Gamma(\Hsm\rightarrow\mathrm{VV})}
  =(a_{1,\RSRadion}+a_{2,\RSRadion})^{2},
  \label{eq:RadionToVV}
\end{equation}
where
\begin{equation}
  a_{i,\RSRadion}
  \deq
    \left\{ \begin{array}{cc}
       a_{i,\RSEigenr} & 
       \quad\begin{array}{c}
           \textrm{if}\,\quad(\RSFundmr>\RSFundmh
           \textrm{ and }\xi^{2}<\frac{\RSFundmh Z^{2}-\RSFundmr}{36\gamma^{2}\RSFundmh}\:)\\
           \,\,\textrm{or\,\,(\,}\RSFundmr\leq\RSFundmh\textrm{ and }\xi^{2}
	   \geq
	   \frac{\RSFundmh Z^{2}-\RSFundmr}{36\gamma^{2}\RSFundmh}\,)
       \end{array}\\
       a_{i,\RSEigenh} & 
       \textrm{otherwise}
    \end{array}\right.\begin{array}{c}
       \\.
    \end{array}
    \label{eq:RadionCouplingCoeff}
\end{equation}
The relative decay width of the Higgs-like state is given by (\ref{eq:RadionToVV})
replacing $a_{i,\RSRadion}$ by $a_{i,\RSHiggs}$, where $a_{i,\RSHiggs}$
is defined accordingly. The following relations for $a_{i,\RSEigenr}$
and $a_{i,\RSEigenh}$ are obtained:
\begin{equation}
  \begin{array}{ccccccc}
     a_{1,\RSEigenr} & 
       \deq & 
       \sin\theta+\frac{6\xi\gamma}{Z}\cos\theta &
       \quad & 
       a_{2,\RSEigenr} &
       \deq &
       \gamma\frac{\cos\theta}{Z}\\
     a_{1,\RSEigenh} &
       \deq &
       \cos\theta-\frac{6\xi\gamma}{Z}\sin\theta &
       \quad &
       a_{2,\RSEigenh} &
       \deq &
       \gamma\frac{\sin\theta}{Z}.
  \end{array}
  \label{eq:RadionCouplingCoeff1}
\end{equation}
Expression~(\ref{eq:RadionToVV}) is valid for all fermions $\mathrm{f}$
and massive vector bosons $\mathrm{V}$ at lowest order. 

In case the Higgs boson or radion is lighter than two times the top
mass, $m_{\mathrm{t}}$, direct decays into top quarks are kinematically
forbidden, but due to the large mass of the top quark, decays into
gluons via top loops are generally not negligible. The matrix element
of a SM Higgs boson decay into gluons is:
\begin{equation}
  \mathrm{ME}(\mathrm{H}_{\mathrm{SM}}\rightarrow\glgl)
    \deq
    \frac{1}{2}\cdot\frac{\alpha_{s}}{8\pi}\cdot\frac{1}{v}\,\Hsm(x)\, 
    F_{\frac{1}{2}}(4m_{t}^{2}/m_{\Hsm}^{2})\,
    \mathrm{G}_{\alpha\mu\nu}(x)\,\mathrm{G_{\alpha}^{\,\mu\nu}}(x).
  \label{eq:HiggsMETopLoop}
\end{equation}
The strong coupling constant is denoted by $\alpha_{s}$, the Higgs
boson mass by $m_{\Hsm}$ and the gluon fields by $\mathrm{G}_{\alpha\mu\nu}$.
The function $F_{\frac{1}{2}}$ is the form factor of the top loop,
which is defined by \cite{RSEinsteinEquation}:
\begin{equation}
  F_{\frac{1}{2}}(\tau)=-2\tau\left[1+(1-\tau)f(\tau)\right],
  \label{eq:TopLoopFormFactor}
\end{equation}
where
\begin{eqnarray}
  f(\tau) &
    = &
    \left\{ \begin{array}{cc}
      \arcsin^{2}\frac{1}{\sqrt{\tau}}, &
        \textrm{if }\tau\geq0\\
      -\frac{1}{4}\left[\ln\frac{1+\sqrt{1-\tau}}{1-\sqrt{1-\tau}}-i\pi\right]^{2}, &
        \textrm{if }\tau<0.
    \end{array}\right.
  \label{eq:TopLoopFormFactorAux}
\end{eqnarray}
 A similar matrix element is obtained for the radion, however it has
the opposite sign and the coupling is reduced by $\gamma$. Since
the radion couples to the trace of the energy momentum tensor, the
anomaly of the trace contributes to the decay width into gluons and
photons in addition to the loop contribution. The anomalous terms
appear in the trace of the renormalised energy momentum tensor in
addition to the unrenormalised trace $\tilde{T}_{\,\mu}^{\mu}$. This
has been shown for example in \cite{TraceAnomaly}. The complete trace
$T_{\,\mu}^{\mu}$ reads:
\begin{equation}
  T_{\,\mu}^{\mu}=
    \tilde{T}_{\,\mu}^{\mu}
    +\frac{\beta}{2g_{\mathrm{R}}}N[F_{\alpha\lambda\rho}F_{\alpha}^{\,\lambda\rho}],
  \label{eq:RenormalisedTrace}
\end{equation}
where $g_{\mathrm{R}}$ denotes the renormalised coupling constant,
$\beta$ the renormalisation group coefficient, $F_{\alpha}^{\,\mu\nu}$
the field strength tensor of strong, electromagnetic and weak interaction
and $N[\ldots]$ normal ordering. Thus, the radion couples directly
to gluon and photon pairs due to the trace anomaly. The additional
coupling to the massive vector bosons is negligible. To fully describe
the coupling of the radion to gluon pairs, the matrix element $\mathrm{ME}(\RSRadion\ra\glgl)$
equivalent of (\ref{eq:HiggsMETopLoop}) has to be extended with the
term:
\begin{equation}
  \mathrm{ME}_{\mathrm{anomaly}}(\RSRadion\ra\glgl)\deq
     \beta\cdot(\alpha_{s}/8\pi)\gamma\RSRadion(x)
     \mathrm{G}_{\alpha\mu\nu}(x)\mathrm{G}_{\alpha}^{\,\mu\nu}(x).
  \label{eq:RadionMETraceAnomaly}
\end{equation}
 For the $\mathrm{SU}(3)$ group of QCD, the renormalisation group
coefficient $\beta$=7. In total, the partial decay width of the radion-like
state becomes \cite{RSEinsteinEquation}:
\begin{equation}
  k_{g}\deq
  \frac{\Gamma(\RSRadion\rightarrow\glgl)}{\Gamma(\mathrm{H}_{S\mathrm{M}}\rightarrow\glgl)}=
  \frac{\left|2\cdot\beta\cdot a_{2,\RSRadion}-(a_{1,\RSRadion}+a_{2,r})
              F_{\frac{1}{2}}(4m_{t}^{2}/m_{r}^{2})\right|^{2}}
       {\left|F_{\frac{1}{2}}(4m_{t}^{2}/m_{r}^{2})\right|^{2}}.
  \label{eq:RadionToGG}
\end{equation}
The factors $a_{i,\RSRadion}$ are those of (\ref{eq:RadionCouplingCoeff}).
The partial decay width of the Higgs-like state, $\Gamma(\RSHiggs\ra\glgl)$
is given by (\ref{eq:RadionToGG}) replacing $a_{i,\RSRadion}$ by
$a_{i,\RSHiggs}$, and $\RSmr$ by $\RSmh$.

\begin{figure}[p]
  \includegraphics{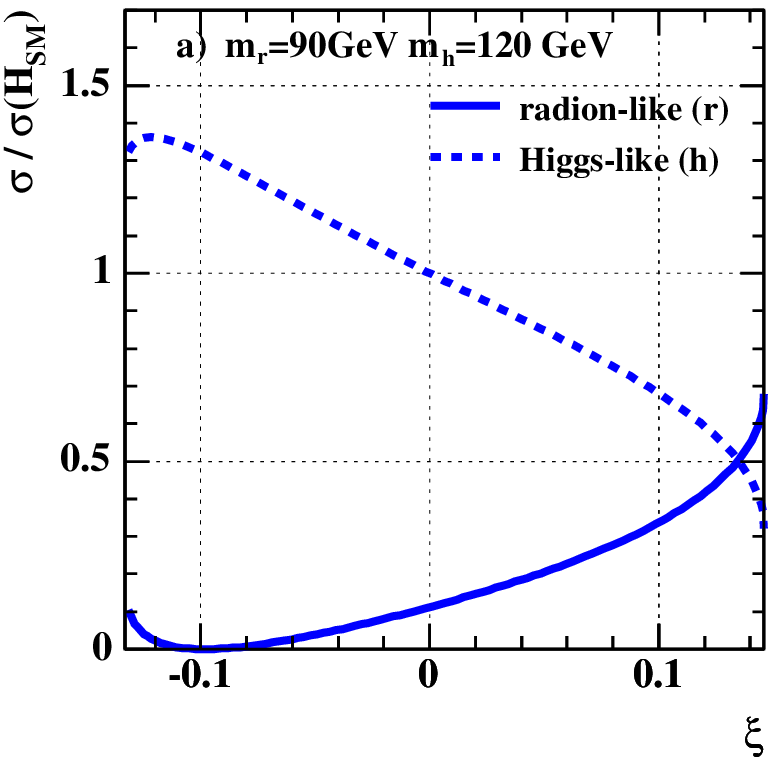}
  \hfill
  \includegraphics{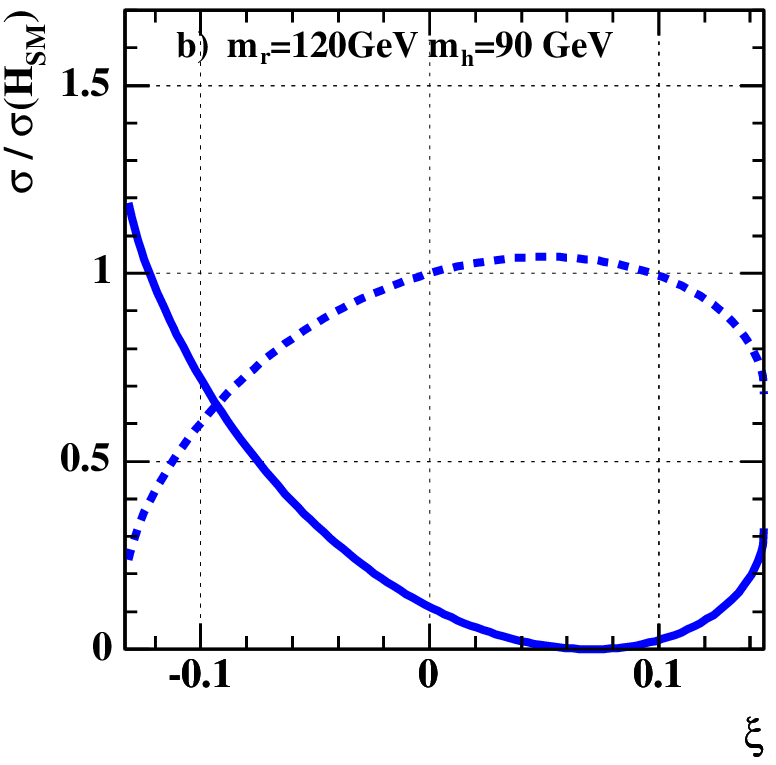}

  \includegraphics{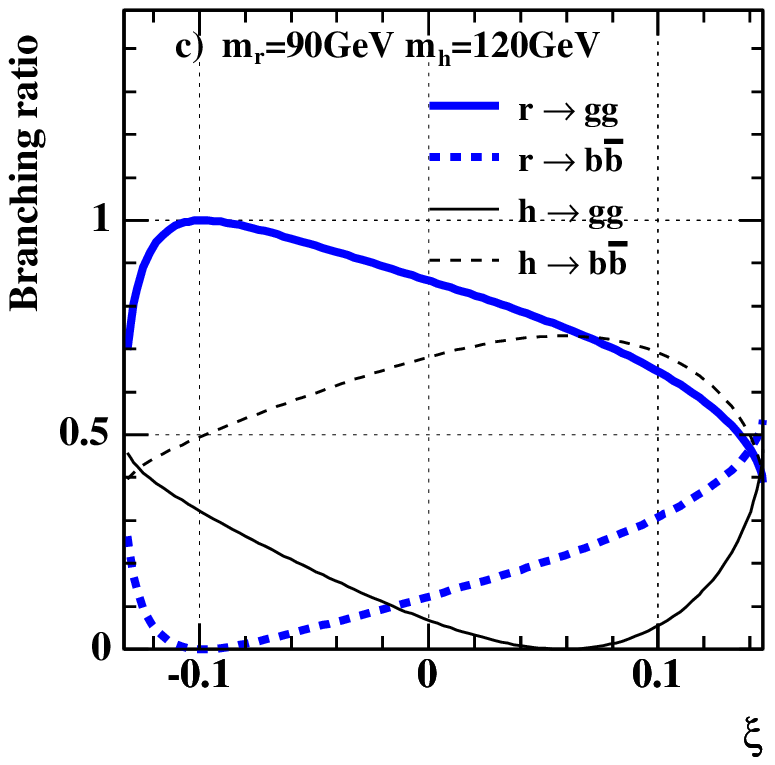}
  \hfill
  \includegraphics{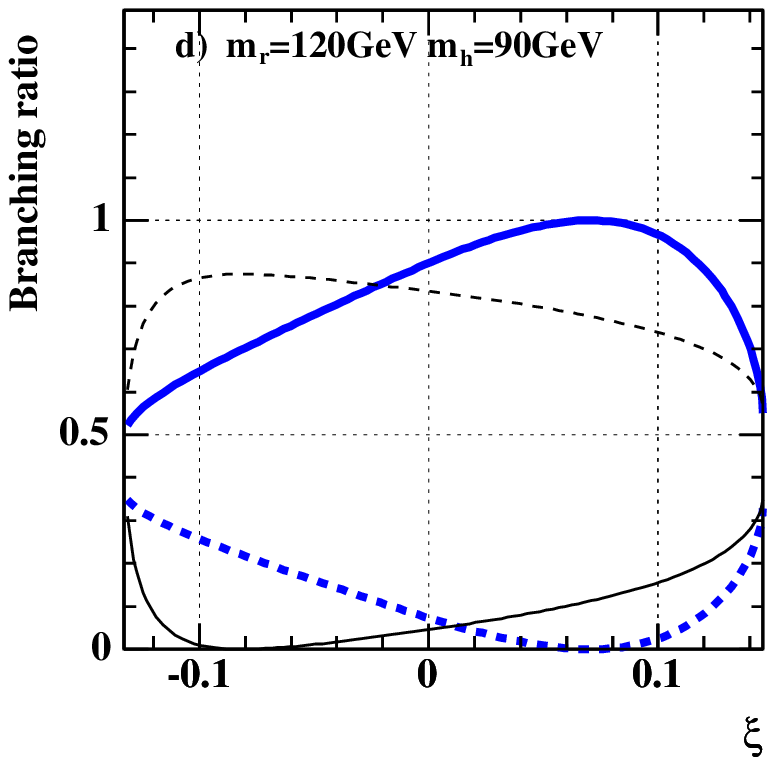}

  \caption{\label{fig:RadionCrossSection}a) and~b) the cross-sections for
     the processes $\ee\ra\Zzero\RSRadion\textrm{ or }\Zzero\RSHiggs$
     of the radion-like and the Higgs-like state, $\RSRadion$ and $\RSHiggs$,
     relative to the corresponding cross-section for a SM Higgs boson for
     two different values of $\RSmr$ and $\RSmh.$ c) and~d) the branching
     ratios of $\RSRadion$ and $\RSHiggs$ into gluon pairs and $\mathrm{b\bar{b}}$.
     The parameter $\Lambda_{\mathrm{W}}$ was chosen to be $300\GeV$.
     The cross-sections and branching ratios of the Higgs-like state $\RSHiggs$
     are identical to those of a SM Higgs boson for $\xi=0$. }
\end{figure}

Except for the additional coupling to gluon pairs and scaled coupling
strength, the couplings of the radion-like and the Higgs-like states
are the same as those of the SM Higgs boson. Thus in $\ee$ collisions
at centre-of-mass energies achieved at LEP, the mass eigenstates,
$\varphi=\RSRadion$ or~$\RSHiggs$, are dominantly produced in the
Higgsstrahlung process, $\ee\ra\mathrm{Z}^{*}\ra\Zzero\varphi$. The
total decay width of the mass eigenstates is smaller than $100\MeV$
for masses of interest ($m_{\varphi}\lesssim115\GeV$). Thus only
decays, $\mathrm{Z}^{*}\rightarrow\mathrm{Z}\varphi$, into on-shell
Higgs bosons or radions have to be considered. The cross-section relative
to Higgsstrahlung in the SM is derived from (\ref{eq:RadionToVV})
and given by:
\begin{equation}
  \frac{\sigma(\ee\rightarrow\mathrm{Z}\varphi)}
       {\sigma(\ee\rightarrow\mathrm{ZH}_{\mathrm{SM}};\,\mHsm=m_{\varphi})}
    =\frac{\Gamma(\varphi\rightarrow\mathrm{VV})}
          {\Gamma(\mathrm{H}_{S\mathrm{M}}\rightarrow\mathrm{VV})}.
  \label{eq:RadionStrahlung}
\end{equation}
In Figure~\ref{fig:RadionCrossSection}, the cross-section and branching
ratios of the two mass eigenstates are displayed as a function of
the mixing parameter $\xi$. Due to the contribution from the trace
anomaly, the radion decays predominantly into a pair of gluons. 

\bibliographystyle{utphys_noeprint}
\bibliography{pr403_slac}


\end{document}